\documentclass[preprint,5p,times,twocolumn]{elsarticle}

\usepackage{amssymb}
\usepackage[numbers]{natbib}
\usepackage{booktabs}
\usepackage{multirow}
\usepackage{subcaption}
\usepackage{afterpage}
\usepackage{longtable}
\usepackage{hyperref}
\usepackage{xcolor}
\journal{Knowledge-Based Systems}

\begin{document}
\let\WriteBookmarks\relax
\def\floatpagepagefraction{1}
\def\textpagefraction{.001}
\begin{frontmatter}



\title{TTMFN: Two-stream Transformer-based Multimodal Fusion Network for Survival Prediction}


\author[1,7]{Ruiquan Ge\corref{cor1}}
\author[1]{Xiangyang Hu}
\author[1]{Rungen Huang}
\author[1]{Gangyong Jia}
\author[2]{Yaqi Wang}
\author[1]{Renshu Gu}
\author[3]{Changmiao Wang}
\author[4]{Elazab Ahmed}
\author[5]{Linyan Wang}
\author[5]{Juan Ye}
\author[6,7]{Ye Li}

\cortext[cor1]{Corresponding author: Ruiquan Ge. E-mail: gespring@hdu.edu.cn.}

\affiliation[1]{organization={School of Computer Science and Technology, Hangzhou Dianzi University},
            city={Hangzhou},
            postcode={310018}, 
            country={China}}
\affiliation[2]{organization={College of Media Engineering, Communication University of Zhejiang},
            city={Hangzhou},
            postcode={310018}, 
            country={China}}
\affiliation[3]{organization={Shenzhen Research Institute of BigData},
            city={Shenzhen},
            postcode={518172}, 
            country={China}}
\affiliation[4]{organization={Misr Higher Institute for Commerce and Computers},
            city={Mansoura},
            country={Egypt}}
\affiliation[5]{organization={Zhejiang University School of Medicine},
            city={Hangzhou},
            postcode={310058}, 
            country={China}}
\affiliation[6]{organization={Shenzhen Institutes of Advanced Technology, Chinese Academy of Sciences},
            city={Shenzhen},
            postcode={518172}, 
            country={China}}
\affiliation[7]{organization={Hangzhou Institute of Advanced Technology},
            city={Hangzhou},
            postcode={310058}, 
            country={China}}

\begin{abstract}
Survival prediction plays a crucial role in assisting clinicians with the development of cancer treatment protocols.  Recent evidence shows that multimodal data can help in the diagnosis of cancer disease and improve survival prediction.
Currently, deep learning-based approaches have experienced increasing success in survival prediction by integrating pathological images and gene expression data.
However, most existing approaches overlook the intra-modality latent information and the complex inter-modality correlations.
Furthermore, existing modalities do not fully exploit the immense representational capabilities of neural networks for feature aggregation and disregard the importance of relationships between features.
Therefore, it is highly recommended to address these issues in order to enhance the prediction performance by proposing a novel deep learning-based method.
We propose a novel framework named Two-stream Transformer-based Multimodal Fusion Network for survival prediction (TTMFN), which integrates pathological images and gene expression data.
In TTMFN, we present a two-stream multimodal co-attention transformer module to take full advantage of the complex relationships between different modalities and the potential connections within the modalities.
Additionally, we develop a multi-head attention pooling approach to effectively aggregate the feature representations of the two modalities.
The experiment results on four datasets from The Cancer Genome Atlas demonstrate that
 TTMFN can achieve the best performance or competitive results compared to the state-of-the-art methods in predicting the overall survival of patients.
\end{abstract}




\end{frontmatter}


\section{Introduction}
\label{}
Cancer is a major global public health issue and a leading cause of death. It is estimated that approximately 20 million new cancer cases emerged, and the number of deaths from cancer reached 10 million in 2020~\citep{RN1}. Accurate prognosis prediction for cancer patients can significantly assist physicians in devising more appropriate treatment plans~\citep{RN2,RN3}. As a result,  survival prediction of human cancers has attracted an increasing attention~\citep{RN4,RN6}.

With the continuous advances in science and technology, various types of data can now be collected and stored, including pathological images, gene expression profiles, copy number variation, DNA methylation, and clinical information, among others~\citep{RN7,RN9}.
This availability of data provides an opportunity to develop numerous biomarkers for cancer prognosis~\citep{RN10,RN11}. 
Pathological images are typically considered the gold standard for diagnosing clinical tumors, as they reveal information on cell morphology~\citep{RN12}.
Cheng et al. extracted hundreds of image features from pathological images,  such as cell size and cell shape, and then analyzed cancer prognosis based on these features, providing valuable insights for further study~\cite{RN13}.
Additionally, gene expression data is widely used to predict the cancer patient's clinical outcome and serve as crucial information for driving cancer prognosis prediction~\citep{RN14}.
Wijethilake et al. identified seven survival-related gene signatures from gene expression data and constructed a prognostic risk model that effectively differentiated between patients with good and poor prognoses~\cite{RN11}.
However, due to the huge data heterogeneity gap between pathological images and gene expression data,  additional challenges arise in prognostic prediction. The size of pathological whole slide images (WSIs) can reach 150000x150000 pixels~\citep{RN15}, and gene expression data contains tens of thousands of features, which significantly exceed the number of samples~\citep{RN16}. 
Furthermore, these data types stem from different collection methods and capture the patient's condition from distinct perspectives. The information they offer is complementary. Consequently, combining pathological data with gene expression data can contribute to achieving improved results in survival prediction.

In recent years, numerous methods based on deep learning have been developed to integrate data from different modalities, which have been successfully employed in predicting cancer prognosis.
For instance, Mobadersany et al. put forward the genomic survival convolutional neural network (GSCNN) for prognosis prediction~\cite{RN17}. The GSCNN combined extracted features from the pathological images with the genomic data features, inputting them into a multi-layer neural network, and subsequently adding non-linear transformations to the features. However, this approach may neglect the intrinsic link between features of different modalities.
Shao et al. initially extracted quantitative image features from pathological images and eigengene module features from gene expression data, then utilized canonical correlation analysis to capture their intrinsic relationships, concurrently identifying critical eigengene module features and image features for survival analysis~\cite{RN18}.
Yao et al. devised a loss function to maximize the correlation between pathological images and gene expression data, proposing the deep correlational survival model~\cite{RN19}.
Following this, Cheerla et al. developed a pan-cancer prognosis prediction architecture, maximizing the loss function based on cosine similarity to enhance the similarity between different modalities~\cite{RN20}.
Wang et al. proposed a bilinear network framework called genomic and pathological deep bilinear network, which was modeled by an intra-modal and inter-modal module for bilinear feature encoding, fully utilizing the intrinsic relationship between different modal features for cancer prognosis prediction~\cite{RN21}. 
Nevertheless, these approaches primarily concentrate on analyzing feature associations and fail to consider heterogeneity between the data of different modalities.
Chen et al. presented a genomic-guided co-attention layer that employed defined genomic embeddings to link the two modalities and reduce redundancy~\cite{RN22}. Gang Wen et al. introduced a novel multi-omics deep survival prediction approach by dually fused graph convolutional network (GCN) referred to as FGCNSurv~\cite{RN49}. However, this approach is heavily dependent on genomic features and directly encodes an extensive amount of patches, which might introduce a significant level of noise. In conclusion, these methodologies encounter several issues including data heterogeneity, disregard of inter-modality feature associations, insufficient data denoising, underutilization of data complementarity, and a lack in modeling inter-modality relationships. Additionally, some researchers have endeavored to address the issue of missing data in multimodal datasets~\cite{RN47}. Furthermore, some individuals have attempted various multimodal methods involving multiple data modalities and have achieved promising results~\cite{RN46,RN48}.

To address these challenges, we introduce a model named TTMFN, designed to improve the accuracy of survival prediction by more effectively leveraging multimodal data. To better exploit the intrinsic relationships within multimodal data, we have developed the TSMCAT module based on the Co-Attention Transformer. Furthermore, we have achieved an advanced level of multimodal feature fusion through the MHAP module, which is founded on the multi-head attention mechanism.

The main contributions of the paper can be summarized as follows:

\begin{itemize}
\item[$\bullet$]we present a method named Two-stream Transformer-based Multimodal Fusion Network for survival prediction (TTMFN), which could better integrates pathological images and gene expression data.
\item[$\bullet$]Our proposed TTMFN comprises a two-stream multimodal co-attention transformer (TSMCAT) module and a multi-head attention pooling (MHAP) module for each modality.
\item[$\bullet$]We develop a feature aggregation strategy based on the multi-head attention mechanism to obtain more complementary information from each modality.
\end{itemize}


\begin{figure*}[ht]
\centering
\includegraphics[width=1\textwidth]{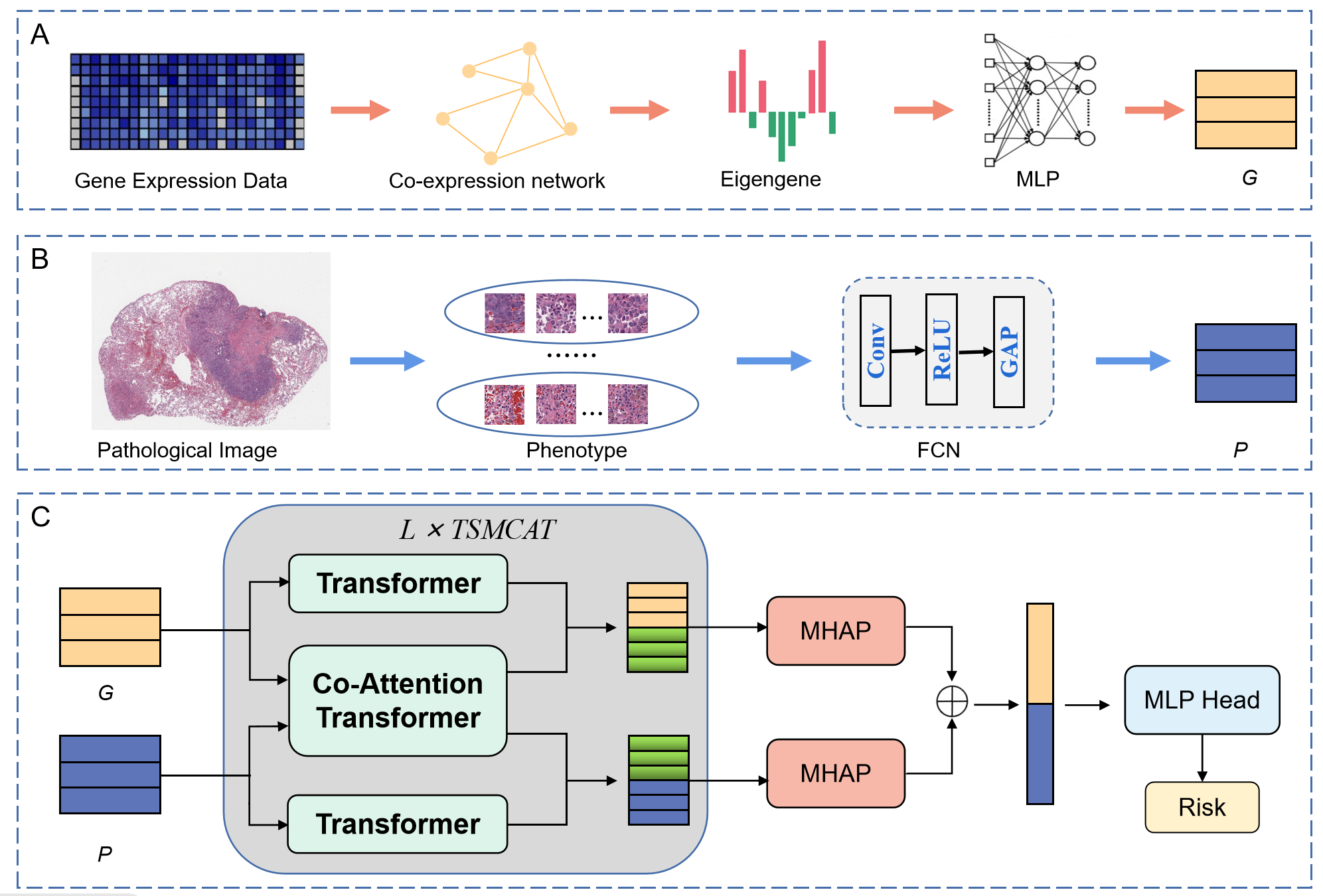}
\caption{The overall framework of the TTMFN, including: (A) gene preprocessing, (B) image preprocessing and (C) the workflow for survival prediction.
}\label{Fig:01}
\end{figure*}

\begin{table}[t]
\vspace{0.1cm}
\renewcommand\tabcolsep{2.0pt}
\caption{\\The numbers of patients, WSIs and gene symbols present in each dataset.}\label{Tab:01}
\begin{tabular*}{\columnwidth}
{@{\extracolsep\fill}ccccc@{\extracolsep\fill}}
\toprule 
Dataset & BRCA & LUAD & BLCA & UCEC\\
\midrule
Number of patients & 957 & 453 & 373 & 444\\
Number of WSIs &1022 & 516 & 437 & 539\\
Number of gene symbols & 20532 &20532 & 16382 & 16382\\
\bottomrule
\end{tabular*}
\end{table}
\section{Materials and methods}
\subsection{Datasets}
In the paper, we primarily utilize four public datasets from The Cancer Genome Atlas (TCGA)~\citep{RN23}: Breast Invasive Carcinoma (BRCA), Lung Adenocarcinoma (LUAD), Bladder Urothelial Carcinoma (BLCA), and Uterine Corpus Endometrial Carcinoma (UCEC). Only matching pathological images and gene expression data are included for each dataset. Gene expression data for TCGA can be obtained from the Broad GDAC Firehose (http://gdac.broadinstitute.org/), a website developed by the Broad Institute. Table~\ref{Tab:01} summarizes the numbers of patients, WSIs, and gene symbols for the two datasets. Consistent with previous studies, we perform 5-fold cross-validation on each dataset. For each fold, 80\% of the data is used for training and validation, while the remaining 20\% is reserved for testing.
\subsection{Overview}
The workflow of the TTMFN is depicted in Fig~\ref{Fig:01}.
Firstly, we extract the modal features from pathological images and gene expression data to construct feature representations for each modality.
Secondly, the extracted feature representation of the two modalities is inputted into the TSMCAT module to exchange features between different modalities.
Thirdly, we employ the MHAP layer to aggregate the fusion features of each modality and then concatenate them to obtain the joint feature representation.
Lastly, the joint feature representation is inputted into the survival prediction layer to obtain the final outcome.
In the following subsection, we will cover each process in greater detail.

\subsubsection{Image feature representation}
Inspired by DeepAttnMISL~\citep{RN24}, our first step involves adopting the OTSU algorithm~\citep{RN25} to extract tissue area from WSIs and sample patches from all WSIs in each patient according to the tissue area.
To capture more detailed information from the image, we sample patches at an objective magnification of 20X (0.5 microns per pixel) and resize them to 512 $\times$ 512 $\times$ 3. Patches containing less than 50\% of tissue sections are discarded.
Next, we extract features for each patch using the network backbone of the pre-trained Resnet18~\citep{RN26} from ImageNet.
After feature extraction, we employ the K-means clustering algorithm to cluster patches into distinct phenotype groups based on the deep learning features. 
These phenotype groups have varying prediction capacities for the patient's clinical result and different numbers of patch samples.
The input can be structured as 1 $\times$ $m_i$ $\times$ \textit{d} for each phenotype, where $m_i$ represents the number of patches in \textit{i}-th phenotype, and \textit{d} is the dimension of deep learning feature.
Finally, we obtain patient-level image feature representation by encoding information of the related phenotype using a simple weight-sharing Fully Convolutional Network (FCN), which can handle varying resolution situations more flexibly. The FCN consists of a 1 $\times$ 1 convolutional layer, a ReLU layer for feature mapping, and an average pooling layer. 
For \textit{i}-th phenotype, its representation is denoted as $p_i\in R^{d_k \times 1}$. Thus, patient-level image feature representation with \textit{C} phenotypes can be defined as $P=[p_1,p_2,...p_C]\in R^{C \times d_k}$.

\subsubsection{Gene feature representation}
To address the challenge of statistical analysis posed by a large number of genes, we implement the gene co-expression network analysis algorithm  to cluster genes into coexpressed gene modules, rather than concentrating on individual genes~\citep{RN27}. Modules are clusters of highly related genes.
To obtain a more expressive representation of embedded features, akin to word embeddings in Natural Language Processing, we treat modules as distinct sets of gene layers with similar representations. This approach differs from traditional algorithms that use singular value decomposition to summarize each module as an eigengene.
Furthermore, to strengthen the connection between genes and images, we set the quantity of modules to be consistent with the quantity of image phenotypes. 
Then, we extract the feature vector of each module using a multilayer perceptron (MLP). For the \textit{j}-th module, its feature representation is denoted as $g_j\in R^{d_k \times 1}$. 
Thus, patient-level gene feature representation can be defined as $G=[g_1,g_2,...g_C]\in R^{C \times d_k}$.

\begin{figure*}
    \centering
    \subfloat[Standard transformer block]{
    \begin{minipage}[c]{0.45\textwidth}
    \centering
    \includegraphics[scale=0.5]{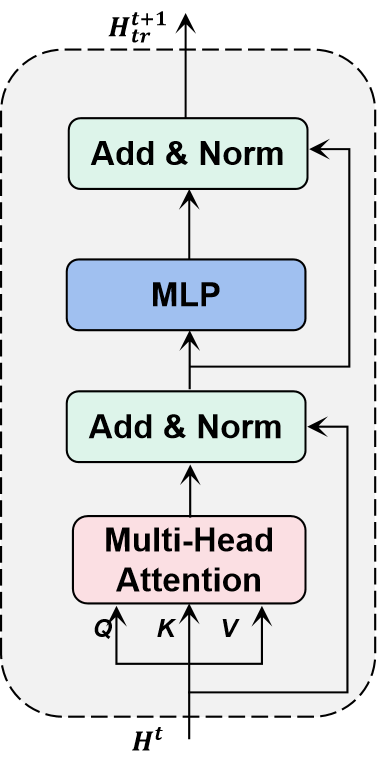}
    \end{minipage}
    }
    \label{Fig tr}
    \subfloat[Co-attention transformer block]{
    \begin{minipage}[c]{0.45\textwidth}
    \centering
    \includegraphics[scale=0.5]{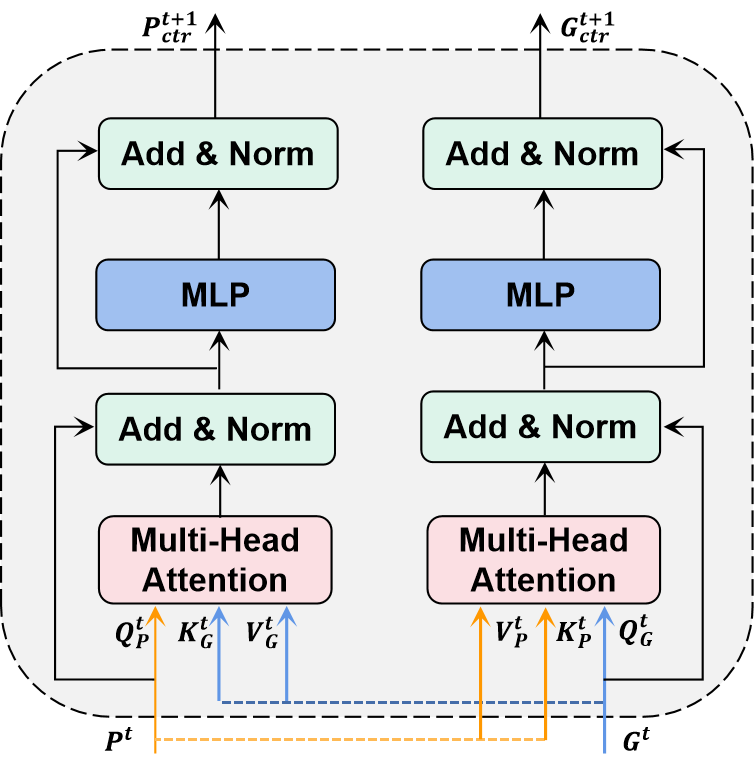}
    
    \end{minipage}
    }
    \label{Fig co-tr}
    
    \caption{The two main blocks of our two-stream multimodal co-attention transformer framework.}
    \label{fig:TSMCAT}
\end{figure*}

\subsubsection{Two-Stream Multimodal Co-Attention Transformer}
In recent years, studies have demonstrated that the accuracy of survival prediction can be enhanced by integrating pathological images with gene expression data.
However, the existing methods fail to utilize feature representations of different modalities for comprehensive multimodal data fusion and overlook the relationships within each modality. 

To address these issues, we propose the TSMCAT module to integrate information from different modalities. As depicted in Fig~\ref{fig:TSMCAT}, the TSMCAT module is based on the Transformer~\citep{RN28}, consisting of two transformer blocks and a co-attention transformer block.
The features of each modality are inputted into the transformer and co-attention transformer, respectively.
We employ the two transformer blocks, as shown in Fig~\ref{fig:TSMCAT}a, to capture intra-modality relationships within gene features and image features, respectively. 
The transformer block comprises an alternating multi-head attention (MHA) block and a MLP block. 
Additionally, a layerNorm layer and residual connections are applied after each MHA block and each MLP block. The MLP contains two fully connected layers and a GELU non-linearity layer. The transformer block can be computed as:
\begin{equation}
  SA = {\rm Attention}(Q,K,V) = {\rm softmax}(QK^T/\sqrt{d})V,
\end{equation}
\begin{equation}
   {\rm MHA}(Q,K,V) = {\rm concat}(SA_1, ... , SA_h),
\end{equation}
\begin{equation}
    H^t_{tr} ={\rm MLP(MHA}(Q_H^{t-1},K_H^{t-1},V_H^{t-1})),
\end{equation}
where $d$ is the query $Q$ dimension and $h$ is the number of self-attention operations.
$H^t_{tr}$ indicates the output features of the $t$-th transformer block. 
Furthermore, in order to comprehensively examine the relationship between various modalities, we introduce the co-attention transformer, as illustrated in Fig~\ref{fig:TSMCAT}b, according to~\citep{RN29}.
A detailed illustration of the co-attention transformer block is provided below.
Given the image and gene feature representation $P^t$ and $G^t$, the co-attention transformer computes query, key, and value matrices, denoted as $Q$, $K$, and $V$, respectively.
We then swap the $K$ and $V$ matrices from each modality and input them into the multi-head self-attention block of the corresponding modality.
In practice, we perform image-conditioned gene attention and gene-conditioned image attention, respectively.
The output of each modality in the $t$-th co-attention transformer is as follows:
\begin{equation}
    P_{ctr}^t ={\rm MLP}({\rm MHA}(Q_P^{t-1},K_G^{t-1},V_G^{t-1})),
\end{equation}
\begin{equation}
    G_{ctr}^t = {\rm MLP}({\rm MHA}(Q_G^{t-1},K_P^{t-1},V_P^{t-1})).
\end{equation}

For each modality, it will be divided into two streams, which are then input to transformer block and co-attention transformer block, respectively. This process yields the intra-modal stream features and inter-modal stream features. Subsequently, we concatenate these features to form the multimodal feature representation for each modality, as shown in the following equations: 
\begin{equation}
    P^{t}= [P_{tr}^t; P_{ctr}^t],
    G^{t}= [G_{tr}^t; G_{ctr}^t].
\end{equation}

\begin{table*}[h]
\centering
\renewcommand\tabcolsep{13.0pt}
\caption{\\Performance comparison of the proposed TTMFN and other methods utilizing C-index value on all datasets.}
\label{tab:all-datasets}

\begin{tabular*}{\textwidth}{@{\extracolsep{\fill}}ccccccc@{\extracolsep{\fill}}}
\toprule
Dataset & Data & Lasso-Cox~\citep{RN31} & EnCox~\citep{RN32} & RSF~\citep{RN33} & GBSA~\citep{RN34} & TTMFN (Ours) \\
\midrule
BLCA & GE & 0.533$\pm$(0.019) & 0.540$\pm$(0.056) & \textbf{0.556$\pm$(0.037)} & 0.556$\pm$(0.039) & 0.555$\pm$(0.044) \\
& PI & 0.612$\pm$(0.014) & 0.601$\pm$(0.059) & \textbf{0.654$\pm$(0.057)} & 0.640$\pm$(0.045) & 0.649$\pm$(0.062) \\
& GE+PI & 0.612$\pm$(0.014) & 0.620$\pm$(0.039) & 0.654$\pm$(0.054) & 0.642$\pm$(0.059) & \textbf{0.661$\pm$(0.067)} \\
\midrule
LUAD & GE & 0.534$\pm$(0.027) & 0.550$\pm$(0.019) & 0.550$\pm$(0.033) & \textbf{0.559$\pm$(0.018)} & 0.553$\pm$(0.044) \\
& PI & 0.652$\pm$(0.047) & 0.619$\pm$(0.031) & 0.667$\pm$(0.052) & 0.669$\pm$(0.050) & \textbf{0.706$\pm$(0.046)} \\
& GE+PI & 0.646$\pm$(0.054) & 0.603$\pm$(0.050) & 0.689$\pm$(0.059) & 0.690$\pm$(0.043) & \textbf{0.727$\pm$(0.038)} \\
\midrule
BRCA & GE & 0.538$\pm$(0.019) & 0.533$\pm$(0.025) & 0.536$\pm$(0.014) & 0.538$\pm$(0.019) & \textbf{0.542$\pm$(0.018)} \\
& PI & 0.612$\pm$(0.014) & 0.651$\pm$(0.014) & 0.654$\pm$(0.022) & 0.647$\pm$(0.022) & \textbf{0.677$\pm$(0.016)} \\
& GE+PI & 0.612$\pm$(0.014) & 0.654$\pm$(0.017) & 0.655$\pm$(0.019) & 0.652$\pm$(0.023) & \textbf{0.697$\pm$(0.020)} \\
\midrule
UCEC & GE & 0.537$\pm$(0.062) & 0.535$\pm$(0.081) & 0.556$\pm$(0.078) & \textbf{0.577$\pm$(0.051)} & 0.398$\pm$(0.043) \\
& PI & 0.616$\pm$(0.057) & 0.611$\pm$(0.029) & 0.639$\pm$(0.089) & \textbf{0.653$\pm$(0.088)} & 0.635$\pm$(0.088) \\
& GE+PI & 0.628$\pm$(0.024) & 0.601$\pm$(0.111) & 0.653$\pm$(0.061) & 0.665$\pm$(0.087) & \textbf{0.671$\pm$(0.062)} \\
\bottomrule
\end{tabular*}
\footnotesize{Note: GE - Gene expression, PI - Pathological images, GE+PI - Gene expression + Pathological images.}
\end{table*}

\subsubsection{Aggregation via Multi-Head Attention Pooling}
\begin{figure}[h]
\centering
\includegraphics[width=0.45\textwidth]{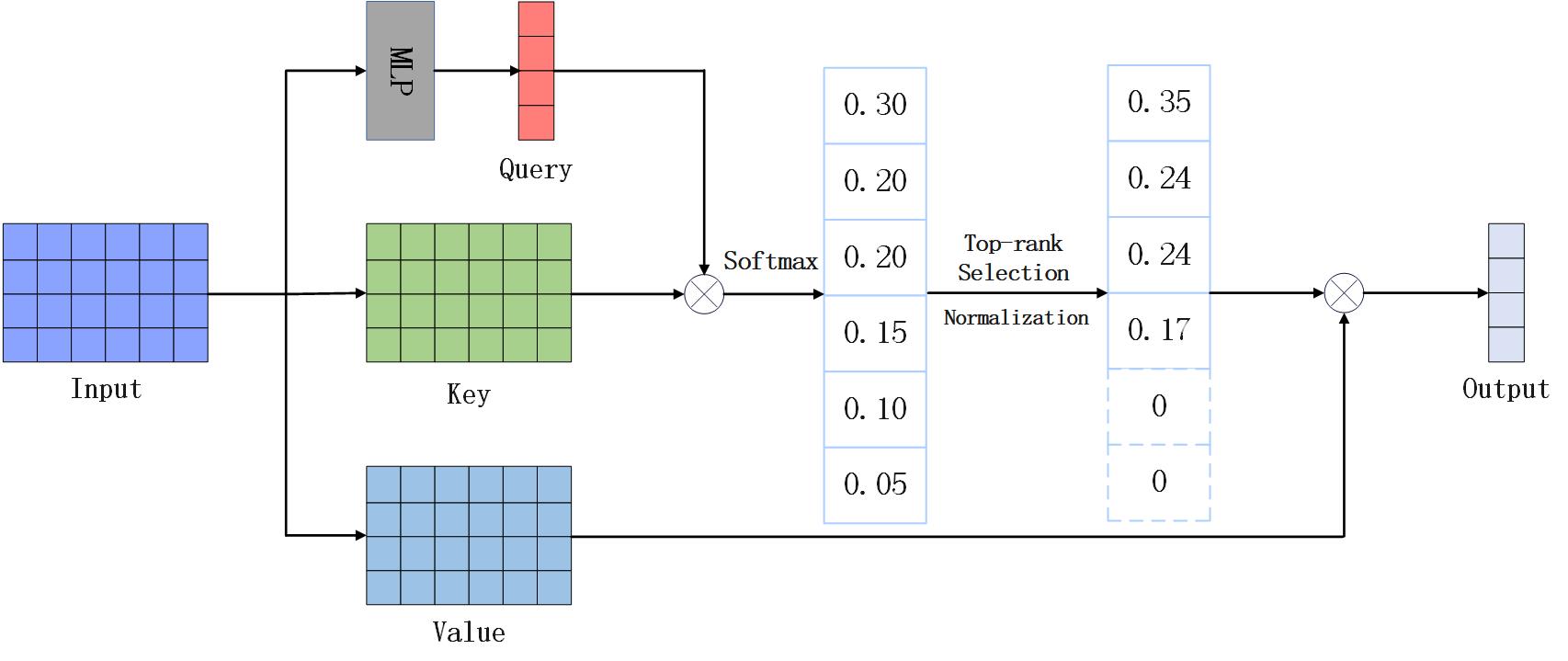}
\caption{ The mechanism of self-attention pooling of MHAP module.}\label{Fig:MHAP}
\end{figure}

Determining the optimal approach to aggregate features from various fusion scenarios into the most efficient patient-level representation is crucial~\citep{RN30}. Common aggregation schemes, such as mean pooling or max pooling, have limitations due to their inflexibility and inability to be tailored for specific tasks. We propose the application of multi-head attention and a network for feature aggregation. In this paper, we present the MHAP mechanism, which consists of multiple self-attention pooling blocks. The self-attention pooling mechanism of the MHAP module is illustrated in Fig~\ref{Fig:MHAP}. Given an input $Z\in R^{ N \times d}$, we can compute the final output as follows:

\begin{equation}
    Q = {\rm MLP}(W^QZ),K = W^KZ,V = W^VZ,
\end{equation}
\begin{equation}
     att = {\rm softmax}(QK^T/ \sqrt{d}),
\end{equation}
\begin{equation}
    {\rm idx = Top-rank}(att, k),  att_{mask} = att_{\rm idx}, 
\end{equation}
\begin{equation}
    y = att_{mask}V,
\end{equation}
where $y$ is the output, $W^Q$, $W^K$, and $W^V \in R^{d \times N}$ are weight matrices. The hyperparameter $k \in$ (0,1] represents the pooling ratio used to determine the number of attention weights to retain. The selected index is denoted as idx, and $att_{mask}$ represents the feature attention mask.
The MLP comprises two layers with a GELU non-linearity, which is employed for dimensionality reduction following the self-attention pooling module. The outputs from each self-attention pooling layer are concentrated to form the final output of the multi-head attention pooling mechanism.

\subsubsection{Survival prediction}
The output obtained after aggregation two modalities, $y_p$ and $y_g$, is concatenated to form the joint feature representation $z$. This representation is then passed through our designed MLP Head module to generate risk score $R$. The MLP Head module comprises an input layer, two hidden layer, and an output layer. To facilitate the nonlinear transformation, we employ ReLU and Sigmoid as the activation functions of the hidden layers and the last output layer, respectively. The module can be outlined as follows:
\begin{equation}
    X = {\rm ReLU}(W_2({\rm ReLU}(W_1 z+b_1))+b_2),
    \label{eq:mlp-risk1}
\end{equation}
\begin{equation}
    R= {\rm Sigmoid}(W_3 X+b_3),
    \label{eq:mlp-risk2}
\end{equation}      
where $W_i$ and $b_i$ refer to weight matrix and bias term of the $i-$th hidden layer, respectively, and R is the output of output layer.

 The negative log partial likelihood~\cite{RN24} is employed as the loss function in our model which is expressed as follows: 
\begin{equation}
    L(R_i) = \sum_{i}\delta _i(-R_i + \log\sum_{j:t_j>=t_i}\exp(R_j)),
    \label{eq:loss}
\end{equation}
where $R_i$ is the $i$-th patient's output, $t_i$, $\delta _i$ are the actual survival time and the censoring status, and $j$ comes from the risk set that the survival time is equal or larger than $t_i$ ($t_j\geq t_i$).

\begin{table*}[h]
\caption{\\Comparative analysis of the C-Index values and AUC values for all methods applied to different datasets.}
\label{Tab:combined}
\begin{subtable}{\textwidth}
\caption{\\BRCA and LUAD datasets.}
\begin{tabular*}{\textwidth}{@{}@{\extracolsep{\fill}}lcccc@{}}
\toprule 
\multirow{2}{*}{Model} & \multicolumn{2}{c}{BRCA} & \multicolumn{2}{c}{LUAD} \\
& C-index & AUC & C-index & AUC \\
\midrule
Attention MIL (WSI Only) & 0.663$\pm$(0.014) & 0.717$\pm$(0.023) & 0.688$\pm$(0.041) & 0.743$\pm$(0.070) \\
Attention MIL (Concat) & 0.665$\pm$(0.015) & 0.724$\pm$(0.019) & 0.698$\pm$(0.061) & 0.740$\pm$(0.078) \\
Attention MIL (Bilinear Pooling) & 0.665$\pm$(0.023) & 0.725$\pm$(0.034) & 0.703$\pm$(0.051) & 0.745$\pm$(0.067) \\
DeepAttnMISL (WSI Only) & 0.650$\pm$(0.018) & 0.737$\pm$(0.022) & 0.691$\pm$(0.050) & 0.689$\pm$(0.081) \\
DeepAttnMISL (Concat) & 0.659$\pm$(0.015) & 0.721$\pm$(0.024) & 0.698$\pm$(0.043) & 0.741$\pm$(0.062) \\
DeepAttnMISL (Bilinear Pooling) & 0.674$\pm$(0.023) & 0.740$\pm$(0.031) & 0.693$\pm$(0.051) & 0.733$\pm$(0.069) \\
MCAT~\citep{RN22} & 0.670$\pm$(0.016) & 0.737$\pm$(0.025) & 0.704$\pm$(0.042) & 0.746$\pm$(0.063) \\
GC-SPLeM~\citep{RN36} & 0.660$\pm$(0.020) & 0.722$\pm$(0.029) & 0.688$\pm$(0.041) & 0.725$\pm$(0.056) \\
\textbf{TTMFN (Ours)} & \textbf{0.697$\pm$(0.020)} & \textbf{0.770$\pm$(0.028)} & \textbf{0.727$\pm$(0.038)} & \textbf{0.770$\pm$(0.052)} \\
\bottomrule
\end{tabular*}
\end{subtable}

\begin{subtable}{\textwidth}
\caption{\\BLCA and UCEC datasets.}
\begin{tabular*}{\textwidth}{@{}@{\extracolsep{\fill}}lcccc@{}}
\toprule 
\multirow{2}{*}{Model} & \multicolumn{2}{c}{BLCA} & \multicolumn{2}{c}{UCEC} \\
& C-index & AUC & C-index & AUC \\
\midrule
Attention MIL (WSI Only) & 0.661$\pm$(0.041) & 0.702$\pm$(0.054) & 0.655$\pm$(0.048) & \textbf{0.742$\pm$(0.053)} \\
Attention MIL (Concat) & \textbf{0.664$\pm$(0.036)} & 0.704$\pm$(0.046) & 0.640$\pm$(0.075) & 0.721$\pm$(0.093) \\
Attention MIL (Bilinear Pooling) & 0.661$\pm$(0.037) & 0.703$\pm$(0.047) & 0.665$\pm$(0.040) & 0.721$\pm$(0.059) \\
DeepAttnMISL (WSI Only) & 0.628$\pm$(0.042) & 0.669$\pm$(0.056) & 0.640$\pm$(0.046) & 0.722$\pm$(0.050) \\
DeepAttnMISL (Concat) & 0.656$\pm$(0.037) & 0.699$\pm$(0.043) & 0.647$\pm$(0.070) & 0.724$\pm$(0.081) \\
DeepAttnMISL (Bilinear Pooling) & 0.657$\pm$(0.041) & 0.698$\pm$(0.051) & 0.652$\pm$(0.041) & 0.722$\pm$(0.064) \\
MCAT~\citep{RN22} & 0.651$\pm$(0.037) & 0.704$\pm$(0.037) & 0.662$\pm$(0.068) & 0.726$\pm$(0.095) \\
GC-SPLeM~\citep{RN36} & 0.644$\pm$(0.032) & 0.694$\pm$(0.038) & 0.647$\pm$(0.064) & 0.719$\pm$(0.062) \\
\textbf{TTMFN (Ours)} & {0.661$\pm$(0.067)} & \textbf{0.707$\pm$(0.066)} & \textbf{0.671$\pm$(0.062)} & {0.734$\pm$(0.098)} \\
\bottomrule
\end{tabular*}
\end{subtable}
\end{table*}
\subsection{Implementation Details}
Our model is implemented using PyTorch library. During the training, we set the learning rate to $1 \times 10^{-4}$ and employ the Adam optimization with weight decay of $5 \times 10^{-4}$.
The pooling ratio $k$ is set to 0.5 for the BRCA dataset and 0.9 for the LUAD dataset.
All hyperparameters are optimized to achieve the best performance on the validation set.
Experiments are conducted using on an NVIDIA TITAN Xp GPU with 12 GB memory.
Due to the varying phenotype sizes of samples, the batch size is set to 1.
Moreover, we monitor the loss of the validation set to prevent overfitting.
If the loss does not improve for 15 consecutive epochs, the training process is early terminated.

\subsection{Evaluation metric}
To assess the performance of our TTMFN, we adopt the concordance index (C-index) as our main evaluation metric. The C-index is commonly utilized to evaluate the performance of survival prediction approaches and can be defined as follows:
\begin{equation}
c = \frac{1}{M}\sum_{i\in \{1...N|\delta _i=1\}}\sum_{s_j \textgreater s_i} I[f_i \textgreater f_j],
\end{equation}
where $M$ denotes the quantity of comparable pairs, and $s$ represents the actual survival time observation. The indicator function is represented by $I[.]$, while $f.$ signifies the corresponding risk. The C-index values range from 0 to 1, with higher values indicating better prediction performance for the model.

\section{Results and discussion}

\subsection{Comparisons with the State-of-the-Art Methods}


\begin{figure*}[h]
    \centering
    \subfloat[The Kaplan–Meier survival curve of different methods on the BRCA dataset]{
    \begin{minipage}[b]{0.45\textwidth} 
    \includegraphics[width=\textwidth]{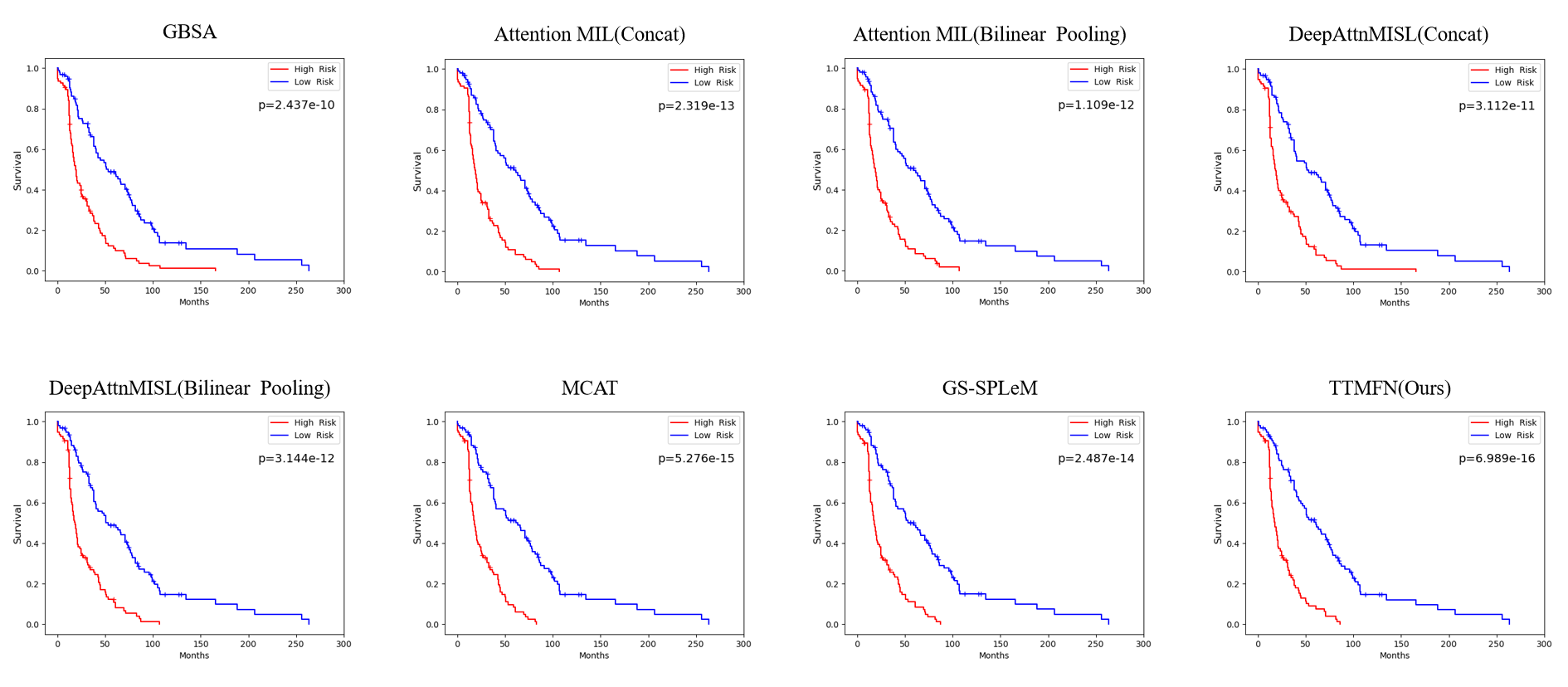}
    \end{minipage}
    }
    \hfill 
    \subfloat[The Kaplan–Meier survival curve of different methods on the LUAD dataset]{
    \begin{minipage}[b]{0.45\textwidth} 
    \includegraphics[width=\textwidth]{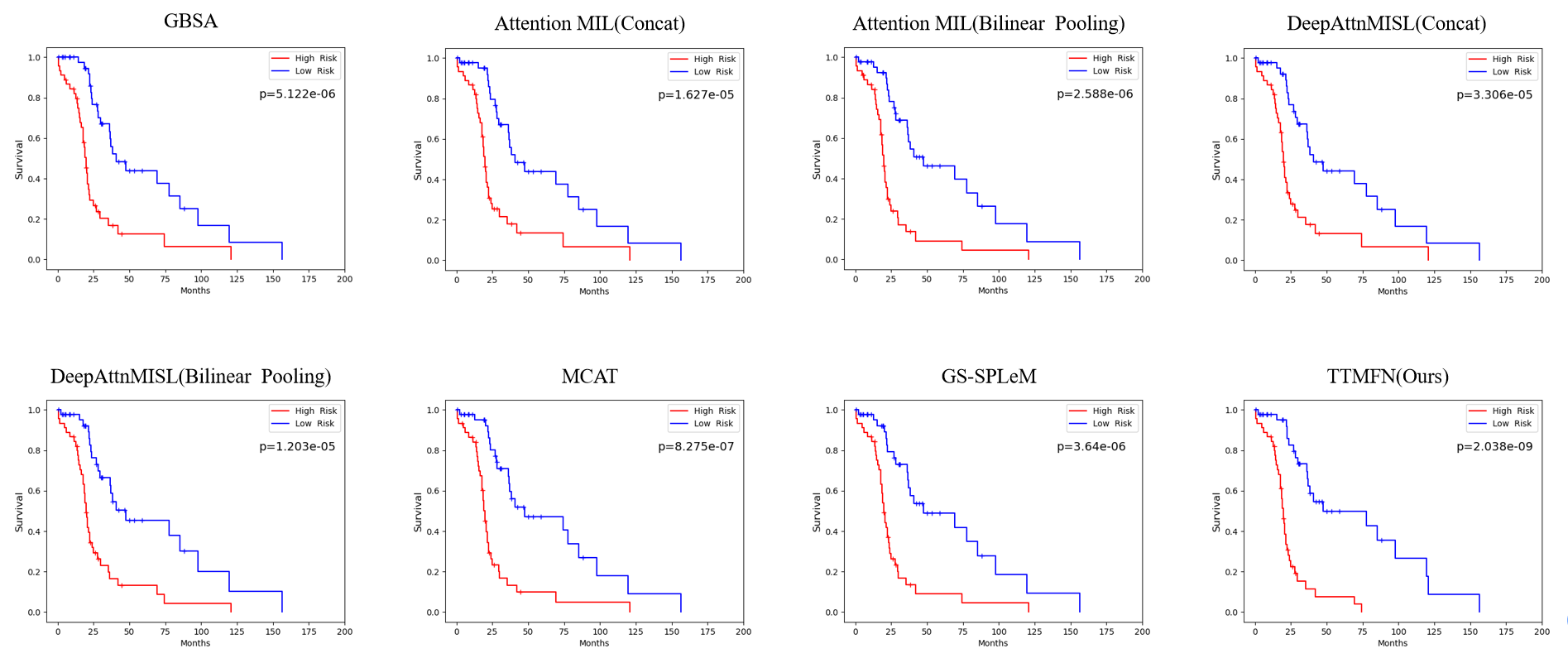}
    \end{minipage}
    }

    \subfloat[The Kaplan–Meier survival curve of different methods on the BLCA dataset]{
    \begin{minipage}[b]{0.45\textwidth} 
    \includegraphics[width=\textwidth]{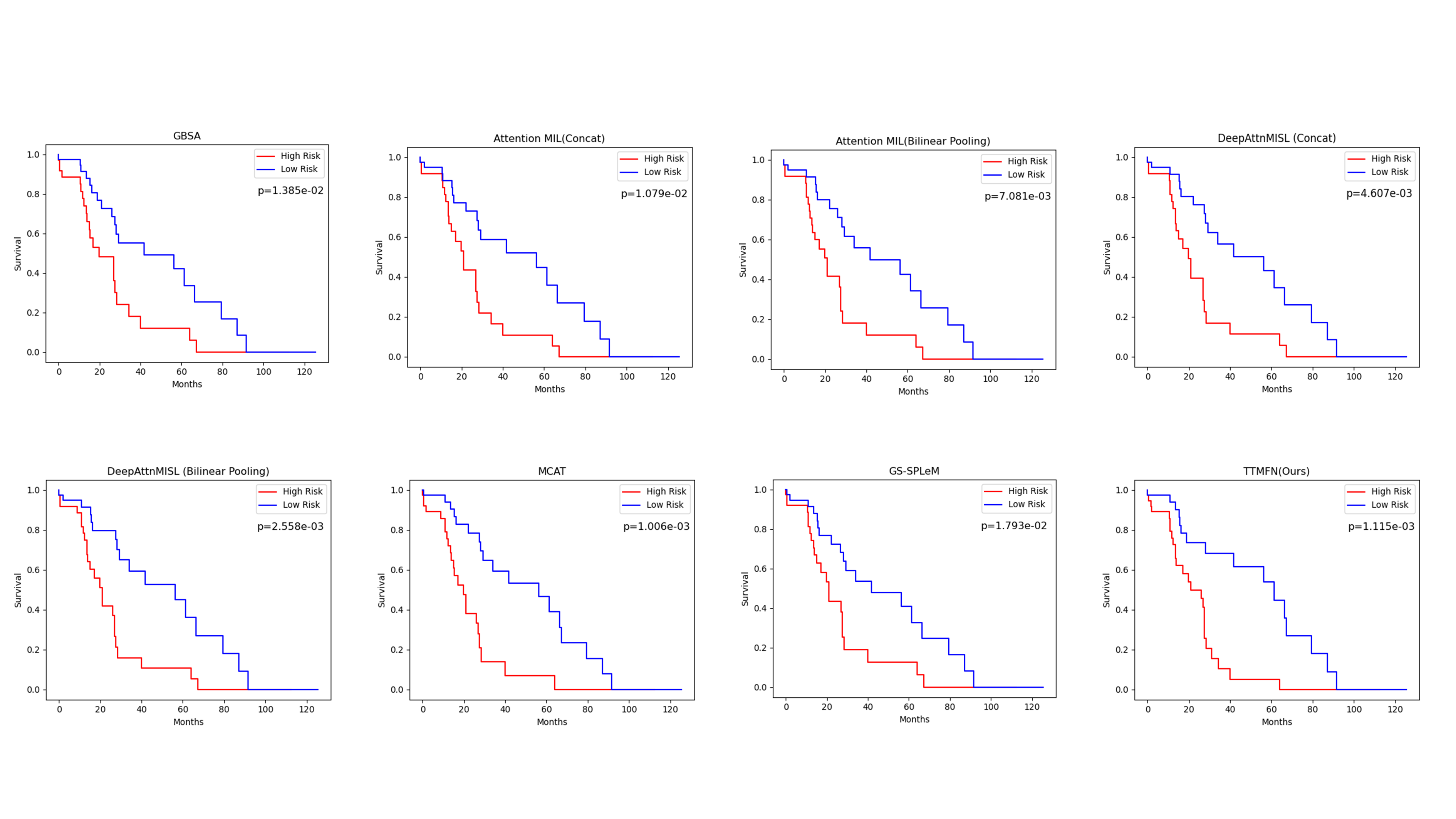}
    \end{minipage}
    }
    \hfill 
    \subfloat[The Kaplan–Meier survival curve of different methods on the UCEC dataset]{
    \begin{minipage}[b]{0.45\textwidth} 
    \includegraphics[width=\textwidth]{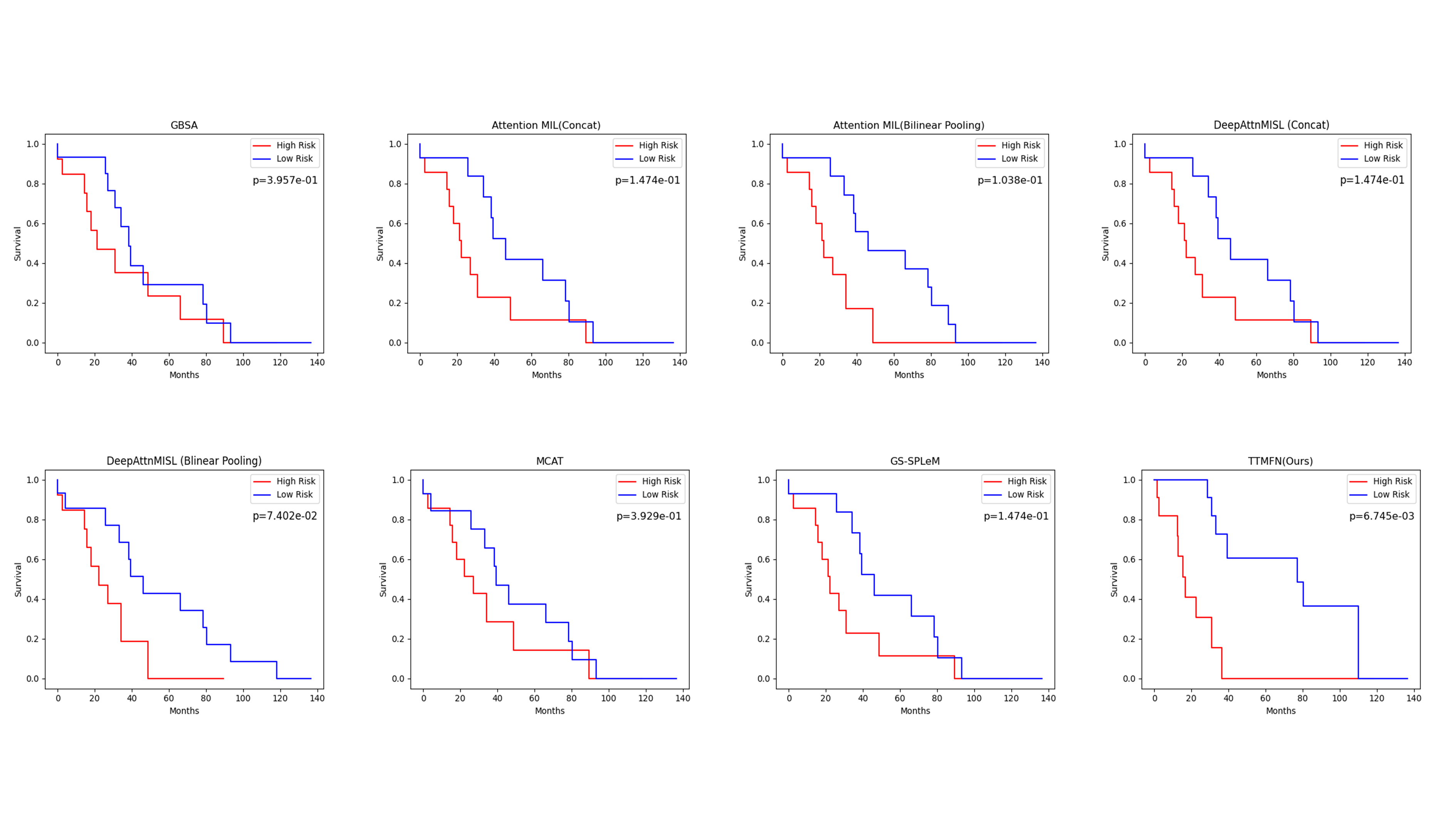}
    \end{minipage}
    }
    \caption{Performance comparison of different models on the all datasets using Kaplan-Meier survival curves. High risk groups are plotted as red lines and low risk groups are plotted as blue lines. The x axis represents the time in months and y axis shows the probability of overall survival. The log rank $p$ value is displayed on each figure.}
    \label{fig:km}
\end{figure*}

To evaluate the performance of our proposed TTMFN, we compare it with several existing machine learning methods, including Lasso-Cox~\citep{RN31}, En-Cox~\citep{RN32}, RSF~\citep{RN33}, GBSA~\citep{RN34}.
For fair comparison, all the compared methods adopt the same features from the gene expression and pathological image data throughout the experiments.
The comparison results for different methods applied to the BRCA, LUAD, BLCA and UCEC datasets are presented in Table~\ref{tab:all-datasets} respectively.
 As evidenced by these tables, integrating pathological images and gene expression data results in higher C-index value for most methods compared to employing single modality data.
For instance, the C-index value of TTMFN using both gene expression data and pathological images increases by 15.5\% and 2\% when compared to the results obtained using only gene expression data and pathological images on the BRCA dataset, respectively. This finding indicates that both gene expression data and pathological images contain valuable predictive information. On the all datasets, the fusion of multimodal methods consistently outperforms single-gene and single-image data, with TTMFN exhibiting strong performance and a C-index improvement of approximately 1\%. Moreover, the use of pathological images alone can yield better performance than using only gene expression data, suggesting that in our method pathological images hold a relatively greater significance than gene expression data for survival prediction.

Subsequently, we compare our TTMFN with several survival prediction methods in computational pathology, employing the same 5-fold cross validation.
The comparison includes the following methods: Attention MIL~\citep{RN35}, DeepAttnMISL~\citep{RN24} MCAT~\citep{RN22}, and GC-SPLeM~\citep{RN36}. 
In addition, we enhance the aforementioned methods with two common fusion mechanisms for multimodal comparisons to TTMFN, specifically concatenation~\citep{RN17} and bilinear pooling~\citep{RN37}.
The C-index values of all methods are presented in Table~\ref{Tab:combined}.
As shown in Table~\ref{Tab:combined}, employing concat and bilinear pooling fusion methods in Attention multiple instance learning (MIL) and DeepAttnMISL results in higher C-index values compared to using only WSIs.

Furthermore, TTMFN consistently achieves the highest C-index values across BRCA, LUAD and UCEC datasets, thereby illustrating its superior performance in survival prediction. Specifically, our proposed TTMFN predominantly obtains the highest C-index value of 0.697 $\pm$ 0.020, 0.727 $\pm$ 0.038 and 0.671 $\pm$ 0.062 on the BRCA dataset, LUAD dataset and UCEC dataset, respectively. This outperforms MCAT by 2.7$\%$, 2.3$\%$, 1.0$\%$ and 0.9$\%$. The outcomes of all datasets are presented in Table~\ref{Tab:combined}. The findings affirm the efficacy of TTMFN in effectively integrating multimodal features, thereby augmenting the precision of survival prediction

Moreover, to further assess the performance of various approaches, we divide the patients into two groups based on the median of predicted risk score.
Patients with risk score less than or equal to the median are considered low risk, while those with risk scores greater than the median are deemed high risk~\citep{RN38}.
In order to assess whether these methods can effectively divide patients into two groups, we plot Kaplan-Meier survival curves~\citep{RN39} for the four datasets and conduct log-rank tests.
The log-rank test is employed to assess the differences between the two curves and determine whether the model is capable of classifying patients into low-risk and high-risk groups.
A decrease in the log-rank test value corresponds to an improvement in the model's survival prediction performance. The Kaplan-Meier survival curves for all methods are displayed in Fig~\ref{fig:km}. As illustrated in Fig~\ref{fig:km}, our TTMFN effectively distinguishes between low-risk and high-risk groups, outperforming other methods. The log-rank test P values for TTMFN on the BRCA and LUAD datasets are 1.054E-32 and 1.910E-22, respectively, which are superior to those of all other methods. In addition, the Kaplan-Meier survival curves for all methods applied to the BLCA data set and UCEC data set are presented in the Fig~\ref{fig:km}, respectively. These curves demonstrate superior results in comparison to other methods. These findings further underscore the effectiveness of TTMFN in survival prediction, suggesting its potential to be applied as a recommendation system for personalized treatment in the future.

\begin{figure*}[htbp]
    \centering
    \includegraphics[width=\textwidth]{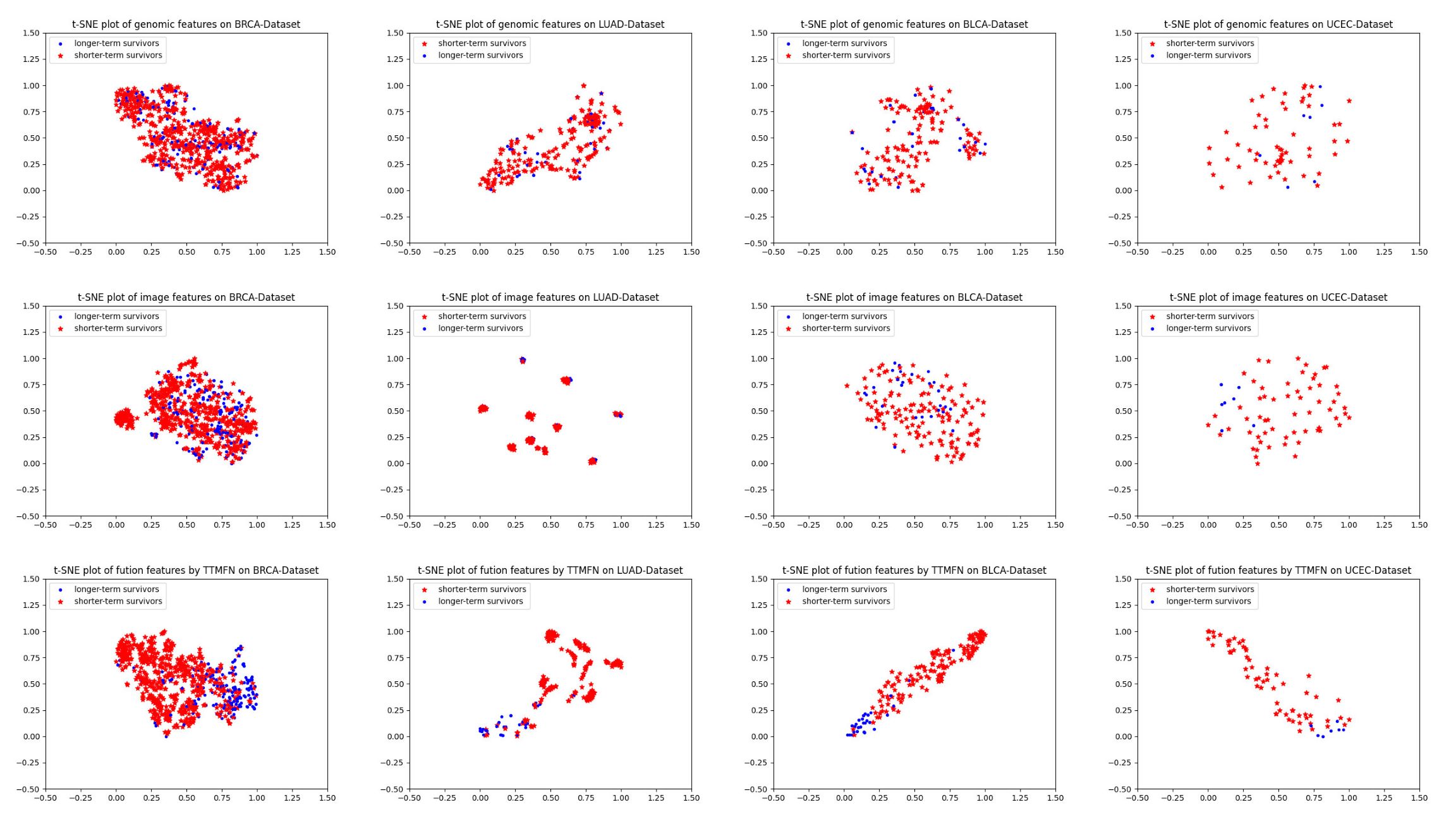}
    \caption{The t-SNE visualization of genomic features, pathological image features and the corresponding fusion features extracted by proposed method on BRCA, LUAD, BLCA, and UCEC Datasets. The red star represents shorter-term survivors, and the blue dot represents longer-term survivors.}
    \label{fig:vi}
\end{figure*}

\subsection{Visualization}
To demonstrate the efficacy of the proposed method in survival prediction, we utilized the widely-used t-SNE algorithm to map the extracted genomic features, image features, and abstract fusion features obtained through TTMFN into a 2-dimensional space, using BRCA, LUAD, BLCA, and UCEC datasets.

As illustrated in Figure ~\ref{fig:vi}, the genomic and pathological image features appear intermingled when visualized concurrently. However, the fusion of these features exhibits distinct patterns, suggesting that integrating data from different modalities can enhance the precision of cancer survival prediction.

Further, it is evident across all BRCA, LUAD, BLCA, and UCEC datasets that the fusion features extracted by TTMFN significantly improve the discrimination between shorter-term and longer-term survivors. This observation implies that our model demonstrates generalization capabilities across different datasets.

\subsection{Ablation Studies}
To further evaluate the effectiveness of the various modules in TTMFN, we conduct an ablation study comparing different model configurations. Firstly, to investigate the effect of the TSMCAT module, we compare the methods that utilized only co-attention transformer from the TSMCAT module with those that used only the transformer to obtain intra-modal features for each modality.
Additionally, we employ meanpool as input into the MLP Head for the experiment. As observed in Table~\ref{tab:123h}, simply using the transformer to extract features and then fusing them does not effectively combine the data from the two modalities. However, by exchanging different modal attention fusions, we achieve a higher C-index value compared to late fusion directly. 
On the BRCA and LUAD datasets, the C-index value of Co-TRM is 0.5$\%$ and 1.8$\%$ higher than that of TRM, respectively. 
Furthermore, our TSMCAT module considers both intra-modal and inter-modal relations, resulting in better C-index value performance than separate considerations.
The C-index value of TSMCAT is 0.8$\%$ and 1.1$\%$ higher than that of Co-TRM on the BRCA and LUAD datasets, respectively. Based on the results provided in Table~\ref{tab:123h}, the TSMCAT module exhibits mediocre  performance in the ablative studies conducted on the BLCA and UCEC datasets.

  

\begin{table*}[t]
\centering
\renewcommand\tabcolsep{22.0pt}
\caption{\\The ablation study of TSMCAT module: Examining the use of the transformer (TRM), co-attention transformer (Co-TRM), and the overall TSMCAT configuration.}\label{tab:123h}

\begin{tabular*}{\textwidth}{@{\extracolsep{\fill}}ccccc@{\extracolsep{\fill}}}
\toprule
         Method & BRCA & LUAD & BLCA & UCEC \\
         \midrule
         TRM & 0.677$\pm$(0.019)& 0.691$\pm$(0.040) & \textbf{0.654$\pm$(0.046)}& 0.644$\pm$(0.064) \\ 
         Co-TRM & 0.682$\pm$(0.013) &0.709$\pm$(0.040) & 0.653$\pm$(0.047)& 0.639$\pm$(0.064)\\
          \textbf{TSMCAT(Ours)} & \textbf{0.690$\pm$(0.012)} & \textbf{0.675$\pm$(0.047)} & 0.644$\pm$(0.040) & \textbf{0.675$\pm$(0.068)} \\
         \bottomrule
    \end{tabular*}

\end{table*}



\begin{table*}[t]
\centering
\renewcommand\tabcolsep{22.0pt}
\caption{\\The comparative study of three Pooling Methods: MeanPool, MaxPool, and MHAP.}\label{tab:123loh}

\begin{tabular*}{\textwidth}{@{\extracolsep{\fill}}ccccc@{\extracolsep{\fill}}}
\toprule
         Method & BRCA & LUAD & BLCA & UCEC \\
         \midrule
         MeanPool &0.690$\pm$(0.012) & 0.720$\pm$(0.047) &0.653$\pm$(0.053) & 0.643$\pm$(0.067)\\
         MaxPool &0.690$\pm$(0.019) & 0.708$\pm$(0.039) &0.643$\pm$(0.033) & 0.650$\pm$(0.068)\\
         MHAP & \textbf{0.697$\pm$(0.020)} & \textbf{0.727$\pm$(0.038)} & \textbf{0.661$\pm$(0.067)} & \textbf{0.671$\pm$(0.062)}\\
         \bottomrule
    \end{tabular*}

\end{table*}

Subsequently, to further assess the performance of the MHAP module in addressing MIL survival prediction tasks, we conduct an ablation study using different pooling methods. Table~\ref{tab:123loh} presents the results for TTMFN employing meanpool, maxpool, and MHAP in image and gene modality feature aggregation. 
Compared to meanpool and maxpool, the performance of our MHAP module improves by 0.7$\%$ and 0.7$\%$ on the BRCA dataset respectively, by 0.7$\%$ and 1.9$\%$ on the LUAD dataset, by 0.8\% and 1.8\% on the BLCA dataset, and by 2.8\% and 2.1\% on the UCEC dataset, respectively.
These results demonstrate that MHAP outperforms traditional meanpool and maxpool in terms of overall C-index value performance. In the MIL survival prediction task, MHAP effectively integrates input information and accurately selects essential features.

\section{Conclusion}
In this study, we propose a novel method, TTMFN, aimed at effectively integrating both pathological images and gene expression data for cancer survival prediction, with the goal of addressing the challenge of intermodal data associations among multimodal data.
The results convincingly demonstrate that survival prediction methods based on multimodal data outperform prediction methods based on single-modality data.
Furthermore, our proposed TSMCAT method exhibits the capability to capture not only the intricate relationship within each modality but also those existing between different modalities. Comprehensive experimental findings underscore TTMFN's superior performance compared to other contemporary methods. Moreover, the Kaplan-Meier survival curve results highlight its robust capacity to distinguish between high-risk and low-risk patients.
Our proposed TTMFN effectively mines the information contained in each modality and seamlessly integrates data from diverse modalities, exhibiting superior performance for survival prediction.
For clinicians, it can potentially facilitate the development of more personalized treatment plans tailored to individual patients.

Despite TTMFN's superior performance in survival prediction, there remains scope for improvement. In recent times, there has been an emergence of multimodal methods employing diverse type of data, a development that we consider to be a promising and potentially fruitful direction. We will seek methods to address the issue of missing multimodal data and attempt to reduce model complexity.
Firstly, to further enhance prediction performance, we will consider incorporating more genomic data such as DNA methylation and copy number variation in the future. 
Secondly, we recognize that certain clinical data may also significantly influence the final prediction result. Thirdly, we plan to implement a graph convolution to facilitate node message passing and a hyperedge mixing mechanism to enable intra-modal and inter-modal interactions between multimodal graphs. Lastly, the application of a deep learning-based preprocessing scheme may further enhance our framework. 
In conclusion, we propose TTMFN, a novel two-stream transformer-based multimodal fusion network for survival prediction, that could be instrumental in guiding prognostic analysis of cancer.

\section*{CRediT Authorship Contribution Statement}
\textbf{Ruiquan Ge:}Conceptualization, Writing - original draft, Project administration.
\textbf{Xiangyang Hu:}Writing - review, editing, Validation, Data curation.
\textbf{Rungen Huang:}Conceptualization, Formal analysis, Writing - original draft, Investigation, Methodology.
\textbf{Gangyong jia:}Project administration, Funding acquisition.
\textbf{Yaqi Wang:}Investigation, Data curation, Investigation, Methodology.
\textbf{Renshu Gu:}Investigation, Writing - original draft.
\textbf{Changmiao wang:}Writing - original draft.
\textbf{Elazab Ahmed:}Writing - original draft.
\textbf{Linyan Wang:}Investigation, Data curation, Funding acquisition.
\textbf{Juan Ye:}Investigation, Writing - original draft, Writing - original draft, Funding acquisition.
\textbf{Ye Li:}Investigation, Writing - original draft.

\section*{Declaration of Competing Interest}

The authors declare that they have no known competing financial interests or personal relationships that could have appeared to influence the work reported in this paper.

\section*{Acknowledgements}
This work has been supported by the Zhejiang Provincial Natural Science Foundation of China (No.LY21F020017,2022C03043), Joint Funds of the Zhejiang Provincial Natural Science Foundation of China (U20A20386), National Natural Science Foundation of China (No. 61702146), GuangDong Basic and Applied Basic Research Foundation (No.2022A1515110570).






\bibliographystyle{elsarticle-num} 
\bibliography{ref}

\end{document}